# Finding Sequential Patterns from Large Sequence Data

Mahdi Esmaeili[1] and Fazekas Gabor[2]

[1] Department of Computer Science, Islamic Azad University (Kashan Branch)
Kashan, Isfahan, Iran

[2] Faculty of Informatics, University of Debrecen
Debrecen, Hungary

**Abstract**
Data mining is the task of discovering interesting patterns from large amounts of data. There are many data mining tasks, such as classification, clustering, association rule mining, and sequential pattern mining. Sequential pattern mining finds sets of data items that occur together frequently in some sequences. Sequential pattern mining, which extracts frequent subsequences from a sequence database, has attracted a great deal of interest during the recent data mining research because it is the basis of many applications, such as: web user analysis, stock trend prediction, DNA sequence analysis, finding language or linguistic patterns from natural language texts, and using the history of symptoms to predict certain kind of disease.
The diversity of the applications may not be possible to apply a single sequential pattern model to all these problems. Each application may require a unique model and solution. A number of research projects were established in recent years to develop meaningful sequential pattern models and efficient algorithms for mining these patterns. In this paper, we theoretically provided a brief overview three types of sequential patterns model.

## 1. Introduction

Sequences are an important kind of data which occur frequently in many fields such as medical, business, financial, customer behavior, educations, security, and other applications. In these applications, the analysis of the data needs to be carried out in different ways to satisfy different application requirements, and it needs to be carried out in an efficient manner [1].
It is obvious that time stamp is an important attribute of each dataset, and it can give us more accurate and useful information and rules. A database consists of sequences of values or events that change with time are called a time series database. This type of database is widely used to store historical data in a diversity of areas. One of the data mining techniques which have been designed for mining time series data is sequential pattern mining.
Sequential pattern mining is trying to find the relationships between occurrences of sequential events for looking for any specific order of the occurrences. In the other words, sequential pattern mining is aiming at finding the frequently occurred sequences to describe the data or predict future data or mining periodical patterns [2][3].
To gain a better understanding of sequential pattern mining problem, let's start by looking at an example. From a shopping store database, we can find frequent sequential purchasing patterns, for example "70% customers who bought the TV typically bought the DVD player and then bought the memory card with certain time gap." It is conceivable that achieving this pattern has great impact to better advertisement and better management of shopping store.
Interestingness measures play an important role in data mining methods, regardless of the kind of patterns being mined. These measures are intended for selecting and ranking patterns according to their potential interest to the user. Choosing interestingness parameters that reflect real human interest remains an open issue.
Sequential pattern mining methods often use the support, which is the criterion to evaluate frequency but this parameter is not efficient to discover some patterns.
Regardless of a great diversity of models for sequential pattern mining, from one perspective, the problem of mining sequential pattern can be partitioned into three categories: periodic patterns, statistically patterns, and approximate patterns [4]. Due to the huge increase in data volume and also quite large search space, efficient solutions for finding patterns in sequence data are nowadays very important. For this reason, using some pruning technique is necessary to optimize the algorithms. They aim to improve performance and reduce response time to find patterns in sequence data.
The remaining of the paper is organized as follows. In section 2 describes the most popular model, periodic patterns. Statistical significant patterns model introduces in section 3. Section 4 presents the flexible model for approximate pattern. Finally, our conclusion is given in section 5.

## 2. Periodic Patterns

This model of pattern is quite rigid and it fails to find patterns whose occurrences are asynchronous [5]. Periodicity detection on time series database is an important data mining task and has broad applications. For example, "The gold price increases every weekend" is a periodic pattern. As mentioned above, this model is often too restrictive since we may fail to detect some interesting pattern if some of its occurrences are misaligned due to inserted noise events. A pattern can be partially filled to enable a more flexible model. For instance, pattern length three $(I_1,*,*)$ is a partial pattern showing that the first symbol must be $I_1$. The system behavior may change over time and some patterns may not be present all the time.

Two parameters, namely *min-rep* and *max-dist*, are used to specify the minimum number of occurrences that is required within each subsequences and the maximum disturbance between any two successive subsequences. The rationale behind this is that a pattern needs to repeat itself at least a certain number of times to demonstrate its significance and periodicity [6].

Couple of algorithms in this area focused on patterns for some pre-specified period length and several models can discover all periodic patterns regardless of the period length. Notice that period usually is apart of what we would like to mine from data. Let us look at an example to explain some definitions.

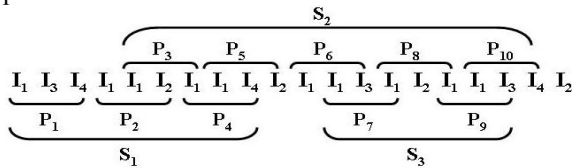

Fig. 1 Example of symbol sequence.

Figure 1 shows matches of partial pattern $(I_1,*,*)$ that is a 1-pattern of period 3. In this Figure, $P_1, P_2,…,$ and $P_{10}$ are ten matches of pattern $(I_1,*,*)$ and $S_1, S_2,$ and $S_3$ are three segments which each one forms a list of N consecutive matches of $(I_1,*,*)$. For example, segment $S_2$ is consist of 5 successive matches of this pattern. $S_1$ and $S_3$ are disjoint segments and $S_1$ and $S_2$ are overlapped segments. In many applications, overlapped segments often are not considered.

If the value of *min-rep* is set to 3, then $S_1$ and $S_2$ qualify as valid segments and $S_3$ is not a valid segment. Two or more than two contiguous and disjoint segments can construct a valid subsequence provided that distance between any two successive valid segments does not exceed the parameter *max-dist*. For example, if the value of *min-rep* and *max-dist* are set to 2 and 3 respectively, both $S_1$ and $S_3$ recognized as valid subsequence whose overall number of repetition is 5.

Given a sequence S, the parameters *min-rep* and *max-dist* and the maximum period length $L_{max}$, in three phases we can discover the valid subsequences that have the most repetitions for each valid pattern whose period length does not exceed $L_{max}$. When parameters are not set properly, noise may be qualified as a pattern. Though, parameter *max-dist* is employed to recognize the noises between segments of perfect repetition of a pattern. The following three phases outline algorithm for mining periodic patterns in brief [4].

The first phase: For each symbol $I$, the distance between any two occurrences of $I$ are examined and then for each period $l$, the set of symbols whose number of times are at least *min-rep* are sent to the next phase. Since there are a huge number of candidates, a pruning method is needed to reduce it.

The second phase: In this phase, the single patterns (1-pattern) are generated. For each period $l$ and each symbol $I$ a candidate pattern $(I,*,*,…,*)$ is formed that number of symbol * is $(l-1)$.

The third phase: After discovering the single patterns in previous phase, i-patterns are generated from the set of valid (i-1)-patterns and then these patterns are validated. In this phase, we can apply some heuristics. For example, it is obvious that if a pattern is valid, then all of its generalizations are valid. Pattern $(I_1,I_2,*)$ is a generalization of pattern $(I_1,I_2,I_3)$.

## 3. Statistically Significant Patterns

The support and confidence are the most popular measures for sequential patterns. The support evaluates frequencies of the patterns and the confidence evaluates frequencies of patterns in the case that sub-patterns are given. These parameters are meaningful and important for some applications. However, in other applications, the number of occurrences (support) may not always represent the significance of a pattern. Sometimes, a large number of occurrences of an expected frequent pattern may not be as interesting as few occurrences of an expected rare pattern. This pattern called surprising pattern instead of frequent pattern. The information gain metric which is widely used in the information theory field, may be useful to evaluate the degree of surprise of the pattern [7]. Target is finding set of patterns that have information gain higher than minimum information gain threshold. Experiments show that the support threshold has to be set very low to discover a small number of patterns with high information gain.

Note that surprising patterns are anti-monotonic. It means advantage of standard pruning techniques such as Apriori property can not be used. For example, the pattern $(I_1,I_2)$ may have enough information gain while neither $(I_1,*)$ nor $(*,I_2)$ does.

Given a pattern P=(I$_1$,I$_2$,…,I$_l$) and an information gain threshold *min-gain*, the goal is to discover all patterns whose information gain in the sequence S exceed the *min-gain* value. Similar to other parameters in data mining algorithms, the appropriate value of the *min-gain* is application dependent and may be defined by a domain expert. There are some heuristics and methods that user can set the value of this threshold.

Information gain of pattern P is defined as follows:

*Info-gain(P)=Info(P)\*Support(P)* (1)

*Info(P)=Info(I$_1$)+ Info(I$_2$)+…+ Info(I$_l$)* (2)

$$Info(I_k) = -\log_{|I|}^{prob(I_k)}$$ (3)

Where *prob(I$_k$)* is probability that symbol I$_k$ occurs and |I| is number of events in S.

For example, the sequence in Figure 1 contains 4 different events I$_1$, I$_2$, I$_3$, and I$_4$. Their probabilities and information gains are shown in Table 1.

Table 1: Probability of occurrence and information gain

| Event | Probability | Information | Information gain |
|---|---|---|---|
| I$_1$ | 10/20=0.50 | -log$_4$(.50)=0.50 | 0.50*10=5.0 |
| I$_2$ | 4/20=0.20 | -log$_4$(.20)=1.16 | 1.16*4=4.64 |
| I$_3$ | 3/20=0.15 | -log$_4$(.15)=1.37 | 1.37*3=4.11 |
| I$_4$ | 3/20=0.15 | -log$_4$(.15)=1.37 | 1.37*3=4.11 |

The main strategy to tackle the problem of mining statistically significant patterns is a recursive method which at the *i*th level of recursion, the patterns with *i* events are examined.

In some applications, users may want to find the *k* most surprising patterns. On the other words, users may want to discover top *k* patterns that their information gain is greater than a threshold. However, the major limitation of information gain value is that it does not recognize location of the occurrences of the patterns.

Let's take a look at two sequences S$_1$ and S$_2$ in Figure 2. The pattern (I$_1$,I$_2$) has the same information gain in the two sequences, it scatter in S$_1$ but repeats consecutive in S$_2$. As a result, it is beneficial if a parameter to evaluate this situation can be specified.

S$_1$ = I$_1$ I$_2$ I$_3$ I$_2$ I$_1$ I$_2$ I$_4$ I$_3$ I$_1$ I$_2$ I$_2$ I$_4$

S$_2$ = I$_2$ I$_3$ I$_4$ I$_2$ I$_1$ I$_2$ I$_1$ I$_2$ I$_1$ I$_2$ I$_4$ I$_3$

Fig. 1 Two examples of symbol sequence.

## 4. Approximate Patterns

Noisy data are commonplace properties of large real world databases. Noise is a random error or variance in a measured variable and there are many possible reasons for noisy data [8]. In many applications, due to the presence of this noise may prevent an occurrence of a pattern and can not be recognized. In addition, a large pattern is much more vulnerable to distortion caused by noise. In these conditions, it is necessary to allow some flexibility in pattern matching. Two previous models only take into account exact match of the pattern in data. An approximate pattern is defined as a sequence of symbols which appears more than a threshold under certain approximation types in a data sequence.

For solving the problem of finding approximate patterns, the concept of compatibility matrix is introduced [4]. This matrix provides a probabilistic connection from observed values to the true values. Based on the compatibility matrix, real support of a pattern can be computed. Table 2 gives an example of the compatibility matrix.

Table 2: An example of compatibility matrix

| True value | Observed value | | | |
|---|---|---|---|---|
| | I$_1$ | I$_2$ | I$_3$ | I$_4$ |
| I$_1$ | 0.80 | 0.15 | 0.00 | 0.05 |
| I$_2$ | 0.10 | 0.70 | 0.10 | 0.10 |
| I$_3$ | 0.0 | 0.00 | 0.90 | 0.10 |
| I$_4$ | 0.10 | 0.15 | 0.00 | 0.75 |

For example, an observed I$_4$ corresponds to a true occurrence of I$_1$ ,I$_2$ ,I$_3$ , and I$_4$ with probability C(I$_1$,I$_4$)=0.05 , C(I$_2$,I$_4$)=0.10 , C(I$_3$,I$_4$)=0.10 , and C(I$_4$,I$_4$)=0.75 ,respectively. Compatibility matrix usually is given by some domain expert but there are some ways to obtain and justify the value of each entry in the matrix so that even with a certain degree of error contained in matrix, sequential pattern mining algorithm can still produce results of reasonable quality.

A new metric, namely match is defined to quantify the significance of a pattern. The combined effect of support and match may need to scan the entire sequence database many times. Similar to other data mining methods, to tackle this problem sampling based algorithms can be used. Consequently, the number of scans through the entire database is minimized.

Given a pattern P=(I$_1$,I$_2$,…,I$_{lp}$) and a symbol sequence S=(I'$_1$,I'$_2$,…,I'$_{ls}$) where *ls≥lp*, match of P in S (denoted by M(P,S)) is defined as the maximal conditional probability P in every distinct subsequence of length *lp* in S. Eq. (4) show how compute match provided that each observed symbol is generated independently.

$M(P,S) = \max_{s \in S} M(P,s)$  if $ls > lp$

$M(P,s) = prob(P|s) = \Pi_{1 \leq i \leq lp} C(I_i, I'_i)$  if $ls = lp$

(4)

For example, the process of calculating the match of two pattern $P_1 = I_2I_1I_3$ and $P_2 = I_2I_3$ in the sequence $S = I_1I_2I_3$ based on compatibility matrix values in Table 2 is shown as follows:

$M(P_1,S) = prob(P_1|S) = \Pi_{1 \leq i \leq 3} C(I_i, I'_i) = C(I_2, I_1) * C(I_1, I_2) * C(I_3, I_3)$

$= 0.1 * 0.15 * 0.9 = 0.0135$

Because of there exist 3 distinct subsequences of length 2 in S ($I_1I_2$, $I_1I_3$, and $I_2I_3$) the match of $P_2$ in S is:

$M(P_2,S) = \max_{s \in S} M(P_2,s) = \max\{M(P_2,I_1I_2), M(P_2,I_1I_3), M(P_2,I_2I_3)\}$

$M(P_2, I_1I_2) = M(I_2I_3, I_1I_2) = C(I_2,I_1) * C(I_3,I_2) = 0.1 * 0 = 0$

$M(P_2, I_1I_3) = M(I_2I_3, I_1I_3) = C(I_2,I_1) * C(I_3,I_3) = 0.1 * 0.9 = 0.09$

$M(P_2, I_2I_3) = M(I_2I_3, I_2I_3) = C(I_2,I_2) * C(I_3,I_3) = 0.7 * 0.9 = 0.63$

$M(P_2,S) = \max\{0, 0.09, 0.63\} = 0.63$

User identifies a minimum match threshold *min-match* to qualify significant patterns. In this model, as the pattern length increases, the match decreases at a much slower pace than the support. For this reason, the number of generated frequent patterns is usually larger than that using the support [9].

## 5. Conclusions

In this work, we provide a brief overview of models of sequential patterns in [4]. The paper theoretically has shown three types of sequential patterns and some properties of them. These models fall into three classes are called periodic pattern, statistically pattern, and approximate pattern. Periodicity can be full periodicity or partial periodicity. In former, every time point contributes to the cyclic behavior of a time series. In contrast, in partial periodicity, some time points contribute to the cyclic behavior of a time series. This model of pattern is so rigid. In many applications, the occurrences of symbols in a sequence may follow a skewed distribution. Using the information gain, as a new metric, help us to discover surprising patterns. To compare the surprising patterns and frequent patterns demonstrates the superiority of surprising patterns. It is clear that the third model, approximate sequential patterns, can provide a powerful means to verify noise. In fact, this is still an active research area with many unsolved challenges. Much more remains to be discovered in this young research field, regarding general concepts, techniques, and applications.

**Mahdi Esmaeili** is currently PhD student at University of Debrecen in Hungary. He received a Bs in software engineering from the University of Isfahan, and his MS in computer science at 1999 in Iran. He is a faculty member of Islamic Azad University (Kashan branch). He has taught in the areas of database and data mining and his research interests include data base, data mining and knowledge discovery, and recent research focusing on the end use of data mining. He has published books about database and information retrieval and also has published articles in international conferences.

**Fazekas Gabor** is an associate professor at the Department of Information Technology of Debrecen University. He received his master degree in mathematics from the Faculty of Sciences of Debrecen University in 1976 and, an awarded PhD degree in 1981. He has taught various topics of computer science and information technology at the Debrecen University for more than 30 years. He was one of the founder of teaching "database management" at Hungarian universities in 1984. His results on code distance and covering radius of codes in polynomial metric spaces achieved a big attention and response from the interested international public. He has supervised many development projects in software engineering, e-learning, coding theory and pattern recognition. His main results have been published in international journals and conferences.